\documentclass[aps,prl,floatfix,superscriptaddress,twocolumn]{revtex4}
\usepackage[latin1]{inputenc}
\usepackage{amsmath}
\usepackage{amsfonts}
\usepackage{amssymb}
\usepackage{graphicx}
\usepackage{subfigure}
\usepackage{pslatex}
\usepackage{verbatim}
\usepackage{color}
\usepackage[sort&compress]{natbib}
\newcommand{\ket}[1]{\left| #1 \right\rangle}

\newcommand{\sitwoeight}{$^{28}$Si}
\newcommand{\sitwonine}{$^{29}$Si}
\newcommand{\mus}{$\rm{\mu}$s}

\begin{document}
\date{\today}
\author{George~C.~Knee~*}
\author{Stephanie Simmons}
\affiliation{Department of Materials, University of Oxford, Parks Road, Oxford OX1 3PH, United Kingdom}
\author{Erik~M.~Gauger}
\affiliation{Department of Materials, University of Oxford, Parks Road, Oxford OX1 3PH, United Kingdom}
\affiliation{Centre for Quantum Technologies, National University of Singapore, 3 Science Drive 2, Singapore 117543}
\author{John~J.~L.~Morton}
\affiliation{Department of Materials, University of Oxford, Parks Road, Oxford OX1 3PH, United Kingdom}
\affiliation{CAESR, Clarendon Laboratory, University of Oxford, Parks Road, Oxford OX1 3PU, United Kingdom}
\author{Helge~Riemann}
\author{Nikolai~V.~Abrosimov}
\affiliation{Leibniz-Institut f\"ur Kristallz\"uchtung, 12489 Berlin, Germany}
\author{Peter~Becker}
\affiliation{PTB Braunschweig, 38116 Braunschweig, Germany}
\author{Hans-Joachim~Pohl}
\affiliation{VITCON Projectconsult GmbH, 07743 Jena, Germany}
\author{Kohei~M.~Itoh}
\affiliation{School of Fundamental Science and Technology, Keio University, Yokohama, Japan}
\author{Mike~L.~W.~Thewalt}
\affiliation{Department of Physics, Simon Fraser University, Burnaby, BC, Canada}
\author{G.~Andrew~D.~Briggs}
\affiliation{Department of Materials, University of Oxford, Parks Road, Oxford OX1 3PH, United Kingdom}
\author{Simon~C.~Benjamin}
\affiliation{Department of Materials, University of Oxford, Parks Road, Oxford OX1 3PH, United Kingdom}
\affiliation{Centre for Quantum Technologies, National University of Singapore, 3 Science Drive 2, Singapore 117543}
\title{Violation of a Leggett-Garg inequality with ideal non-invasive measurements}
\begin{abstract}
\noindent
The quantum superposition principle states that an entity can exist in two different states simultaneously, counter to our `classical' intuition. Is it possible to understand a given system's behaviour without such a concept? A test designed by Leggett and Garg can rule out this possibility. The test, originally intended for macroscopic objects, has been implemented in various systems. However to-date no experiment has employed the `ideal negative result' measurements that are required for the most robust test. Here we introduce a general protocol for these special measurements using an ancillary system which acts as a local measuring device but which need not be perfectly prepared. We report an experimental realisation using spin-bearing phosphorus impurities in silicon. The results demonstrate the necessity of a non-classical picture for this class of microscopic system. Our procedure can be applied to systems of any size, whether individually controlled or in a spatial ensemble.
\end{abstract}
\maketitle
\noindent
There is a stark contrast between the way we think of the microscopic world (which is well described by quantum physics) and the way we experience the everyday macroscopic world (which appears to follow rules which are  altogether more intuitive). There have been a number of proposals for experimental tests which pit quantum physics against alternative views of reality: for example the theorems of Bell~\cite{Bell1964} and of Kochen and Specker~\cite{KochenSpecker1967}. Corresponding laboratory tests have been performed and to-date support the necessity of quantum physics. But even if a quantum description of the microscopic world is necessary, we face the equally profound question of understanding the relationship between the quantum world and our familiar classical experience. Some thinkers, such as Penrose, suggest that there are as-yet undiscovered physical laws which prevent superposition of `macroscopic' states~\cite{Penrose1996}. Most physicists would agree that sufficiently large objects (such as the moon) must indeed ``be there" when nobody looks. The Leggett-Garg inequality~\cite{LeggettGarg1985} was developed in order to address this question. The protocol may be applied to systems of arbitrary size, thus theories which hold that quantum theory breaks down at some particular scale can be experimentally tested. 

Limited variants of the Leggett and Garg (LG) test have been reported for microscopic objects such as photons~\cite{DresselBroadbentHowell2011,GogginAlmeidaBarbieri2011} or nuclear spins~\cite{WaldherrNeumannHuelga2011} and for the larger superconducting `transmon' system~\cite{Palacios-LMalletNguyen2010}. The approach presented here represents the first implementation of LG's powerful `ideal negative result' measurement procedure. We describe a general protocol for such measurements, introducing an ancillary system~\cite{PazMahler1993} which acts as a local measuring device. Importantly we can account for imperfect preparation of the measuring device through a quantity which we call `venality'. We find that at some finite venality (typically corresponding to a thermal threshold) the LG test becomes possible. Our procedure can be employed for any physical system where a suitable ancilla can be adequately initialised; it thus provides a test for a system of any size, whether addressed as part of a spatial ensemble or controlled individually.

For a given system with two suitably defined states, our protocol provides the opportunity to invalidate the conjunction of the following two beliefs:
\emph{Macrorealism} (MR) - the system is always in one of its macroscopically distinguishable states; and
\emph{Non-invasive measurability} (NIM) - it is possible in principle to determine the state of the system without altering its subsequent evolution.
A quantum physicist will typically reject NIM, but crucially the test requires only that the macrorealist accept it~\cite{Peres1988,LeggettGarg1989}. In a test of the above assumptions, a compelling argument for the non-invasiveness of the measurements should be made in a language acceptable to a macrorealist. Leggett-Garg inequality violations that have been reported with weak measurements \cite{Palacios-LMalletNguyen2010,DresselBroadbentHowell2011,GogginAlmeidaBarbieri2011} employ a measurement procedure which may ultimately fail to convince a macrorealist that the measurements are indeed non-invasive. Proposals for experimentally determining the invasiveness of each measurement exist~\cite{WildeMizel2011}, but we make use of Leggett and Garg's arguments for the non-invasiveness of an `ideal negative result' measurement scheme. Other experiments have been performed~\cite{Palacios-LMalletNguyen2010,WaldherrNeumannHuelga2011} which use the assumption of `stationarity'~\cite{JordanKorotkovButtiker2006,RuskovKorotkovMizel2006,WilliamsJordan2008}. This assumption severely narrows the class of macrorealist theories which are put to the test  (please see Supplementary Methods); we do not make this assumption and so our method tests a wider class of theories.

We employ a method which equips a two level system with a local measuring device: another two-level system~\cite{PazMahler1993}. We refer to the system being tested as the `primary system' and the associated measuring device as the `ancilla'. We consider how macrorealists might approach an imperfectly prepared measuring device, showing that even an `adversarial' macrorealist who makes the most extreme assumptions about the effects of invasive measurements must nevertheless expect certain constraints. Quantum physics predicts that under certain conditions such constraints can still be violated. We show that although the primary system may be in a totally mixed state, the degree to which the ancilla is correctly initialised directly affects one's ability to violate the constraint. We implement our protocol experimentally using an ensemble of nucleus-electron spin pairs in phosphorus doped silicon. The results comprehensively rule out a large range of classical descriptions for this class of system, which although microscopic represents an important step towards performing rigorous tests on more macroscopic systems.
\section{Results}
\subsection{Three core experiments}
Consider the primary system's two states of interest labelled by $\uparrow$ or by $\downarrow$ undergoing arbitrary dynamics governed by a process labelled $U$. If the system is probed at distinct times with a measurement which distinguishes one state from the other (Figure \ref{sixcircuits}a), the degree to which the state of the system correlates with itself at the different times may be quantified. The two-time correlator $K_{ij}=\langle Q(t_i)Q(t_j)\rangle$ is the expected value of the product of the measurement outcome of the observable $Q$ at time $t_i$ and at time $t_j$. If  $Q\in \{+1,-1\}$ for $\uparrow,\downarrow$ respectively, 
and since the correlator is an average, we have $-1\leq K_{ij}\leq 1$. Calculating this quantity is straightforward: one simply measures at $t_i$, waits, and measures again at $t_j$ multiplying the results together to compute $Q(t_i)Q(t_j)$. One then averages over many instances of the experiment either by repeating it many times, or by employing an array of many identical systems, as in a recent test of non-contextuality \cite{MoussaRyanCory2010}. Although in a spatial ensemble one has no access to individual elements, because of the ancillary nature of the measuring qubit (each element of the ensemble is coupled to its own), the test may still be performed. 

Now consider a family of three experiments, each one beginning with a primary system in an identical initial state $\rho_s$ and evolving under identical conditions governing the dynamics of the state. In the first experiment measurements are made at $t_1$ and $t_2$ to determine $K_{12}$. In the same way the second and third experiments are used to determine $K_{23}$ and $K_{13}$ (Figure \ref{sixcircuits}b). We then evaluate the `Leggett-Garg Function' \cite{LeggettGarg1985}:
\begin{align}
f=K_{12}+K_{23}+K_{13}+1.
\end{align}
Any macrorealist theory according to which the measurements $Q$ are non-invasive must predict $f\geq0$. This is true regardless of how the theory distributes probability arbitrarily amongst classical trajectories of the primary system (the assumption of `Induction' is required, see Ref. \cite{Leggett2008}, Supplementary Methods). In contrast, according to quantum physics, $f$ is negative for suitably chosen time evolution operator $U$. 
\subsection{Ideal negative result measurements}
Following Leggett \cite{LeggettGarg1985,Leggett1988,Leggett2002,Leggett2008}, we implement measurements of $Q$ which, by exploiting MR, are `extremely natural and plausible'~\cite{LeggettGarg1985} candidates for non invasiveness.
Imagine a measuring device that is physically incapable of interacting with a system in state $\uparrow$, but that will (possibly invasively) detect a system in state $\downarrow$. Suppose we apply this detector to our system and it does not `click'. The macrorealist infers the system is in state $\uparrow$, and was in this state immediately prior to measurement -- but this information is obtained without any interaction. Switching to a complementary measuring device that perceives only the $\uparrow$ state allows one to obtain the full set of data non-invasively, as long as one always abandons all experiments where the detector clicks. 

One must acknowledge that it is impossible to ensure that the measurement apparatus does not couple to and disturb some other, hidden, degrees of freedom. One cannot exclude macrorealist theories involving interactions between hidden parts of the system and detector (which in our case would have to occur even during a null measurement event). This is a general point applying to any LG test: one can only address a subclass of macrorealist theories which hold that such irremediable hidden degrees of freedom either do not exist, or are not relevant.

\begin{figure}[t]
\includegraphics[width=8.5cm]{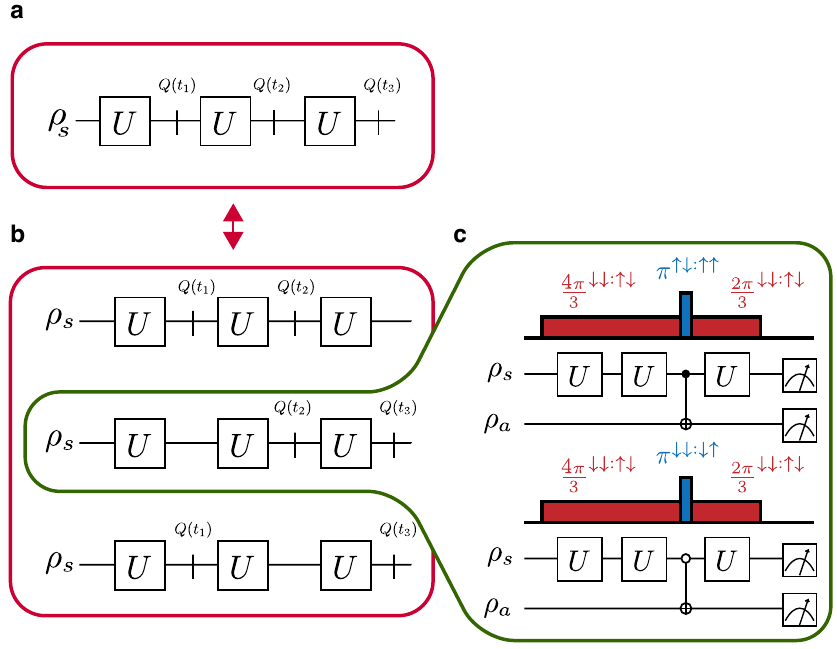}
\caption{\label{sixcircuits}\textbf{Our full implementation of the LG test requires six sub-experiments.} If the measurements are non-invasive, the outcome statistics of \textbf{a,} a single ideal experiment (where all measurements are made in each run) will match those of \textbf{b,} a set of three core experiments (where only two measurements are made in each run). The actual lab implementation for the second of the three core experiments is shown in panel \textbf{c}. Shown in colour are the corresponding pulses applied to our experimental coupled-spin$\frac{1}{2}$ system. The primary system is driven with radio-frequency pulses (red areas), and the {\sc cnot} and anti-{\sc cnot} operations are each applied with a single selective microwave frequency pulse (blue areas). The other two core experiments are similarly resolved into a pair of complimentary sub-experiments.} 
\label{3exps}
\end{figure}

The use of two detector configurations means that the three experiments introduced previously are each further resolved into a pair of experiments, one for non-invasive measurement of $\uparrow$, and one for $\downarrow$ (Figure \ref{sixcircuits}c).  We utilise either a {\sc cnot} gate (which will flip the state of the ancilla if the control, i.e. the primary system, is in $\downarrow$) or use an anti-{\sc cnot} gate (which will flip the state of the ancilla qubit if the primary is in $\uparrow$; Figure~\ref{sixcircuits}), in each case post selecting experimental runs where the gate was not triggered (Supplementary Methods). The second, final measurement in each experiment need not be implemented non-invasively, since the subsequent dynamics are irrelevant. Note that it is important that the physical implementation of the {\sc cnot} (and anti-{\sc cnot}) operation is such that the primary system receives no perturbation when it is in the state associated with a null result.

Here we set $U=\cos\frac{\theta}{2}\mathbb{I}+i\sin\frac{\theta}{2}\sigma_x$. 
As long as the ancilla is correctly initialised, the quantum prediction is $K_{ij}=\cos(\theta)$ independent of $\rho_s$ and hence
\begin{align}
f=2\cos\theta+\cos2\theta+1,
\end{align}
which takes the value $f=-0.5$ for $\theta=2\pi/3$, violating the inequality $f\geq 0$ predicted under MR $\cap$ NIM. Arguments constraining the macrorealist to non-negative values for $f$ also do not depend on the primary system's initial state.

\subsection{Corrupt ancillas}
For any protocol employing a measurement ancilla, its initialisation is of fundamental importance. A macrorealist regards an imperfectly prepared primary-ancilla qubit pair as a statistical mixture of the four states $\ket{\downarrow\downarrow},\ket{\downarrow\uparrow},\ket{\uparrow\downarrow},\ket{\uparrow\uparrow}$ and similarly a quantum physicist describes the initial state as a density matrix diagonal in the $|{\rm system}\rangle|{\rm ancilla}\rangle$ basis. Quantum mechanically an incorrectly initialised ancilla will give rise to an incorrect correlator sign. To the macrorealist it will give a false indication that the measurement had been noninvasive, allowing a potentially corrupt element through the postselection. We define the venality $\zeta$ as the fraction of the ensemble for which the ancilla is incorrectly prepared. Quantum physics predicts that each $K_{ij}$ generalises to $(1-\zeta)K_{ij}-\zeta K_{ij}$, leading to 
\begin{align}
f\rightarrow (1-2\zeta)(2\cos\theta+\cos2\theta)+1.
\end{align}
We identify two macrorealist attitudes pertaining to the effect of an invasive measurement. A `moderate' view is that any invasively perturbed systems act in a random way, and so average to produce zero net correlation. Then $K_{ij}\rightarrow (1-\zeta)K_{ij}$ and so with $g=K_{12}+K_{23}+K_{13}$ and $g\geq -1$ for a macrorealist,
\begin{align}
f^{\textrm{moderate}}=(1-\zeta)g+1\geq \zeta.
\end{align}
Note $f$ is still constrained to be non-negative. An `adversarial' view is that invasively perturbed elements will, by some unidentified process, act in such a manner as to minimise $f$. Consequently  $K_{ij}\rightarrow (1-\zeta)K_{ij}-\zeta$  so that
\begin{align}
 f^{\textrm{adversarial}}=(1-\zeta)g-3\zeta+1\geq-2\zeta.
\label{advinq}
\end{align}
This is the most aggressive stance available to a macrorealist. 
\begin{figure}[t]
\includegraphics[width=8.5cm]{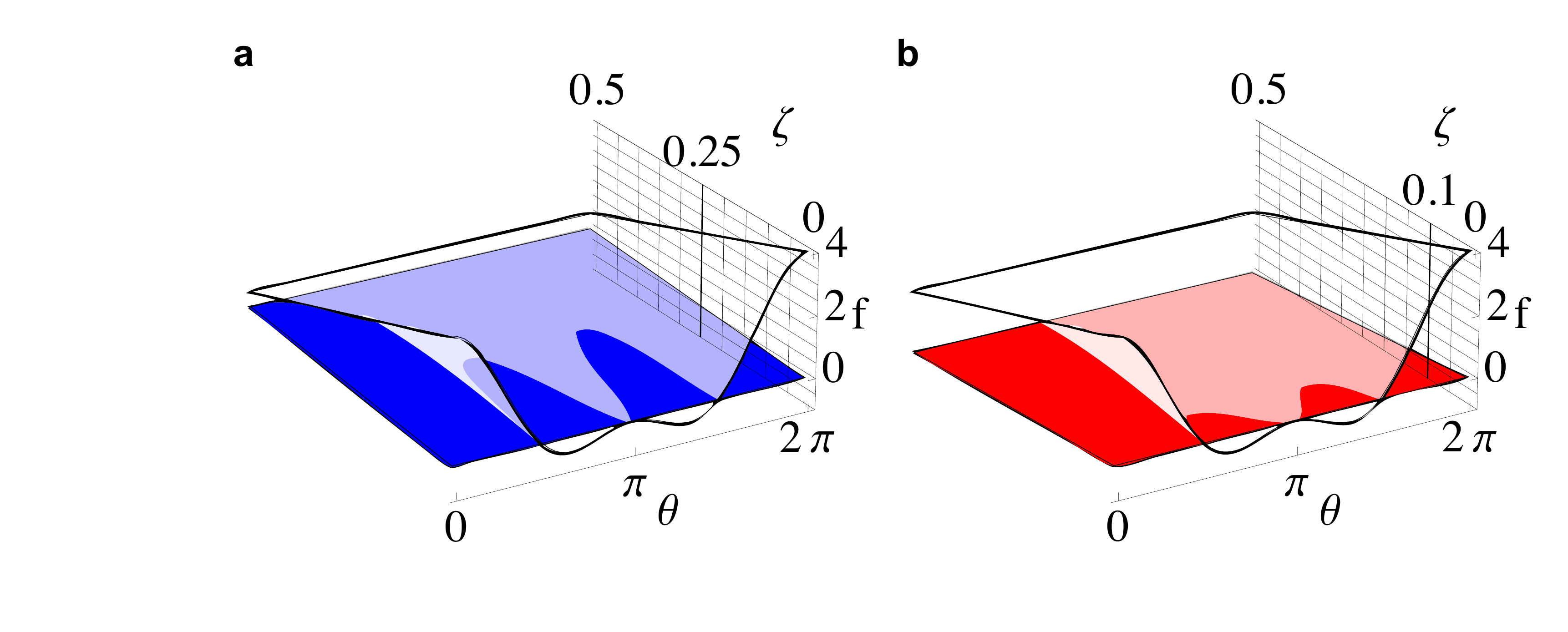}
\caption{\label{redandblack}\textbf{The bounds on the LG inequality for quantum mechanical and macrorealist models depend on the venality in the experiment.} Plots of the quantum mechanical prediction (white) and lower bound of a modified inequality for the
\textbf{a}, moderate (blue)  
and  \textbf{b}, adversarial (red) macrorealist attitudes as a function of the angle $\theta$ and the venality $\zeta$. Where the quantum prediction dips below the macrorealist bound it is in principle possible to invalidate the macrorealist stance. Note the critical value of $\zeta=0.25$ and $\zeta=0.1$ above which one cannot exclude macrorealism for the moderate and adversarial approaches respectively.}
\end{figure}
The relevant thresholds are plotted in Figure \ref{redandblack}, showing that minimising $\zeta$ is crucial for a successful experiment. 

\subsection{Experimental implementation}
To demonstrate an experimental violation of these inequalities, we consider an ensemble of phosphorus donors in silicon, consisting of electron-nuclear spin pairs. Here the nuclear spin is the primary system, while the electron is the measurement ancilla. 
In the high field limit, the eigenstates of this spin $\frac{1}{2}$ -- spin $\frac{1}{2}$ system are precisely the four product spin states. In thermal equilibrium, and ignoring the weak polarisation of the nucleus, these states are populated according to the Boltzmann distribution, where the spin states are in the ratio $\alpha:1$ for $\alpha=\textrm{exp}(-g\mu B/k_B T)$. Here $B=3.357$~T is the magnetic field, $g$ is the electron spin's $g$-factor, $\mu$ is the Bohr magneton, $k_B$ is Boltzmann's constant and $T$ is the temperature.
\begin{figure}[t]
\begin{center}
\includegraphics[width=8.5cm]{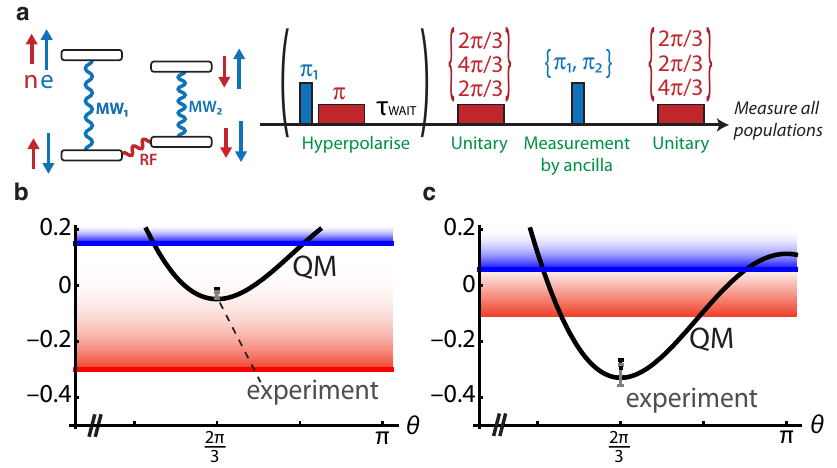}
\caption{\label{datapoints}\textbf{Experimental values for the LG function are compared with bounds from quantum mechanics and macrorealist theories} \textbf{a}, The populations of the four system-ancilla (nucleus-electron) states are manipulated with microwave and radio-frequency radiation.  The experimentally determined value of the Leggett-Garg function at a static field of $B=3.357$ T is plotted \textbf{b} at 2.6~K for  a thermal initial state  and \textbf{c} at  2.7~K with a hyperpolarised initial state. The minimum bound for each macrorealist approach is also plotted: blue for moderate, red for adversarial. Error bars represent uncertainty in measurement of the final state, and the grey point and error bars are the result of correcting for known measurement errors (namely the population damping effects of the tomography pulse sequence).}
\end{center}
\end{figure}
The electron and nuclear spin are coupled through a 117.5 MHz hyperfine interaction, which distinguishes each individual $\ket{\uparrow}$ : $\ket{\downarrow}$ transition. The electronic (nuclear) transitions can be individually addressed using selective microwave (radio-frequency) pulses. The unitary nuclear rotation $U$ may be performed in a manner which is conditional on the system being in the `correct' ancilla state $\downarrow$ (as a refinement of the circuit illustrated in Figure \ref{sixcircuits}c) because the postselected data will always correspond to the unitary operation $U$ having been applied. The correlator sequences applied to this system are shown in Figure \ref{datapoints}a. The final measurement at the end of an individual correlator sequence is accomplished through population tomography \cite{SimmonsBrownRiemann2010}.
\subsection{Inequality violation}
We performed two experimental tests with results shown in Figure \ref{datapoints}b and \ref{datapoints}c. The first used a simple state in thermal equilibrium at $2.6$~K with $\zeta=2\alpha/(2+2\alpha)=0.150$, yielding $f=-0.031$. The second used an established hyperpolarisation sequence~\cite{SimmonsBrownRiemann2010} from an initial state at $2.7$~K. Due to the conditional nature of $U$ this technique reduces the venality (please see Supplementary Methods) to $\zeta=2\alpha^2/(1+\alpha+2\alpha^2)=0.056$, yielding $f=-0.296$.   In the course of our experiments, the fidelity of the final state populations with respect to the ideal target was never less than $98.9\%$. Our analysis has made two assumptions about the measurement process: Firstly, that any detector imperfections do not conspire to favour anticorrelations preferentially. Secondly, as discussed earlier, that our null measurements do not influence the correlations through some hidden structure of the macrorealist's state. Our results then constitute a falsification of MR $\cap$ NIM for cold nuclear spins.
\section{Discussion}
Our approach relies upon the `ideal negative result' measurements originally envisaged by LG; we show that such measurements are possible through an ancilla. Recognising that ancilla preparation will always be imperfect, we account for the implications through a quantity termed `venality'. We show that for sufficiently low venality even an `adversarial' macrorealist must concede that his view is inconsistent with experimental results. Importantly this approach allows one to employ either individually controlled systems or a spatial ensemble, and it is applicable to systems of any size.

For our chosen experimental system, an ensemble of phosphorous impurities in silicon, we were able to reach a low temperature, high field regime where the venality is low enough for our LG test to be feasible. Through the use of high precision control techniques, we were indeed able to obtain a result representing an unequivocal violation of the inequality.  The violation of this bound has secured the following profound conclusion: All accurate descriptions of systems of this type must include a concept similar to that of quantum superposition, and/or an exotic notion of measurement similar to that of wavefunction collapse.

While our experimental results relate to a microscopic system, we emphasise that our protocol is entirely general in terms of the scale of the system and whether it is individually controlled. Thus we hope that our work will give rise to a series of experiments which probe successively more macroscopic entities with the same rigour that we apply here. Ultimately such experiments will realise Leggett and Garg's vision of establishing whether superpositions of macroscopically distinct states are indeed possible. 

\section{Methods}
\subsection{Weak measurements versus ideal negative result measurements}
LG tests employ the concept of non-invasive measurement in a fundamental way; the approaches one may take when seeking an implementation include weak measurement or ideal negative result measurement. Weak measurements are likely to be regarded by both the quantum physicist and the macrorealist as \emph{approximations} to true non-invasiveness. Meanwhile Leggett's concept of negative result measurement will seem \emph{highly} invasive to a quantum physicist but entirely non-invasive to a macrorealist. As we are interested in a test involving a gap between the predictions of quantum physics versus macrorealist theories, it is the latter approach that is preferable.
The weak measurement approach cannot be altered to take account of the amount of invasiveness by defining something like the venality (which is a measure of how often a non-ideal measurement is applied and not a measure of the invasiveness of a given measurement). A back action is imparted for each and every run of the experiment, and so the so called `clumsiness loophole' \cite{WildeMizel2011} cannot be closed this way. 

\subsection{Sample preparation}
Si:P consists of an electron spin $S=1/2$ ($g = 1.9987$) coupled to the nuclear spin $I=1/2$ of $^{31}$P through an isotropic hyperfine coupling of $a=4.19$~mT. The W-band EPR signal comprises of two lines (one for each nuclear spin projection $M_I = \pm 1/2$). Our experiments were performed on the low-field line of the EPR doublet corresponding to $M_I=1/2$. At 2.6~K and 3.36~T, the electron and nuclear spin $T_1$~were measured to be approximately 1~s and 100~s, respectively. 

The sample consists of a \sitwoeight-enriched single crystal about 0.5~mm in diameter with a residual \sitwonine~concentration of order 70 ppm, produced by decomposing isotopically enriched silane in a recirculating reactor to produce poly-Si rods, followed by floating zone crystallisation. Phosphorus doping of $\sim10^{14}$ cm$^{-3}$ was achieved by adding dilute PH$_3$ gas to the Ar ambient during the final float zone single crystal growth.  Further information on the sample growth has been reported elsewhere~\cite{BeckerPohlRiemann2010}.

Pulsed EPR experiments were performed using a W-band (94~GHz) Bruker Elexsys 680 spectrometer equipped with a 6T superconducting magnet and a low temperature helium-flow cryostat (Oxford CF935). The cryostat was pumped to achieve a temperature of 2.6~K (internal thermocouple). Typical pulse times were 56~ns (288~ns) for a MW1 (MW2) $\pi$ pulse and 90~$\mu$s for an RF $\pi$ pulse.
\subsection{Spin resonance experiments}
Both the conditional nuclear operation, and also the non-invasiveness of the measurement operation performed by the ancilla electron spin, require that the magnetic resonance pulses are selective to a high degree. The electron and nuclear spin resonance frequencies are separated by $\sim10$ and $\sim10^4$ times the pulse excitation bandwidth respectively, so we may rule out excitation of non-resonant spin transitions (please see Supplementary Methods). The spin-relaxation lifetimes at 2.6~K are orders of magnitude longer than the total experiment time of 450~\mus, and so we expect (and observe) no population shifts due to relaxation on these timescales. 

The Leggett-Garg function $f$ is a linear combination of populations, which can be considered as diagonal entries in a density matrix. Using magnetic resonance, only population differences can be measured. This leads to an `observable' (or `pseudopure') component which can be manipulated by an experimentalist, and an `unobservable' component, made up of populations common to all eigenstates. For each of the six sub-experiments, a four dimensional `pseudopure' matrix was measured, which was then added to an appropriately scaled identity component determined by the local magnetic field and temperature of the sample (representing the unmeasurable component of the ensemble). A baseline measurement was taken as an average of 2000 samples, and all data sets were baseline-corrected before processing. The population differences were measured by an average of 200 samples and scaled with respect to a measured thermal amplitude (also taken as an average over 200 samples), and adjusted to have unit trace with the addition of an appropriately scaled identity matrix. 
\subsection{Error analysis}
The errors corresponding to each population were calculated according to the standard error of the direct difference measurements. These population errors were transformed into final Leggett-Garg function uncertainty by a Monte Carlo generation of density matrices. The generated matrices deviated from the measured matrix in each element by an amount chosen randomly from a normal distribution whose standard deviation matched that elements' error. Once re-normalised, unphysical matrices were discarded and statistics on physical matrices were collected. In total, $2^{12}$ matrices were used to compile the final uncertainty. This constituted the `raw' pseudopure matrix.

The principal source of error in the population difference measurements came from microwave and radio-frequency inhomogeneity leading to a spread in applied rotation angles across the ensemble. These errors constituted a loss of signal for every applied pulse, with a negligible net over- or under-rotation. We fit the Rabi oscillations of each of the two microwave-frequency rotations and the radio-frequency rotations to arrive at an estimate for the signal lost per applied $\pi$ rotation in the population tomography sequence. These fits were used to estimate the populations without the amplitude-dampening effects of the tomography sequence, and the uncertainties of these fits were used to estimate the uncertainty of each population element. These uncertainties were combined with the measurement uncertainty error before performing Monte-Carlo simulations as above with $2^{12}$ matrices. This enables us to correct for the limitations of the tomography sequence and infer the actual populations before the tomography is applied. 

The calculated pseudopure matrix $\rho_{\textrm{pp}}$ was added to the appropriate amount of identity matrix $\mathbb{I}$ as determined by the sample temperature. The explicit reconstruction is given by
\begin{align}
\rho_F=[ {\alpha}/(2(1+{\alpha}))] \mathbb{I} + [(1-{\alpha})/((1+{\alpha}))]\rho_{\textrm{pp}}\nonumber.
\end{align}
The diagonal entries of six matrices of this kind were used to generate each of the datapoints shown in Figure \ref{datapoints}.  The value for $f$ calculated from raw populations is shown there in black and the value for $f$ calculated from populations corrected to compensate for the principal tomography errors is shown in grey, for both the hyperpolarised and un-hyperpolarised data sets. 

\begin{figure}[h]
\includegraphics[width=8.5cm]{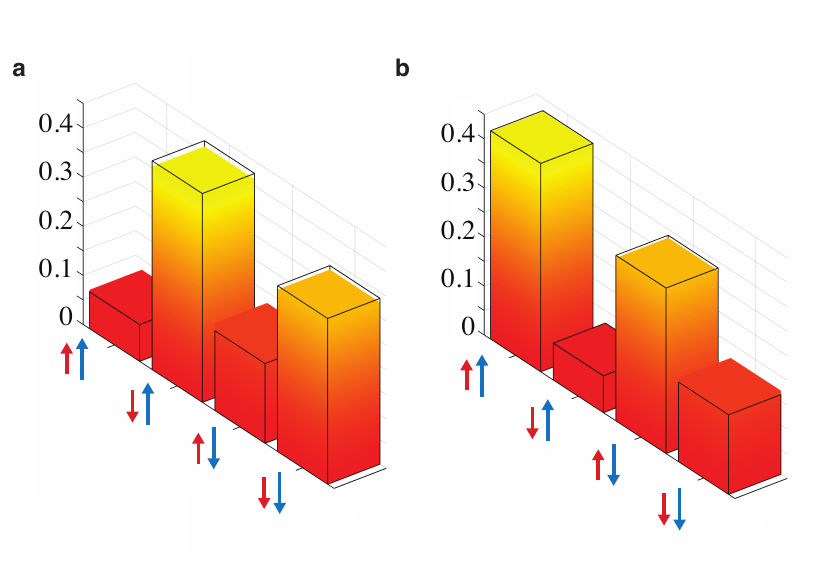}
\caption{\label{skyscrapers}\textbf{An example of the measured populations acquired from tomography.} Orange bars represent diagonal matrix elements at the end of the second core experiment. The wireframes are the ideal quantum values. The populations were acquired from \textbf{a}, the {\sc cnot} circuit and \textbf{b}, anti-{\sc cnot} circuit.}
\end{figure}

There are two conventional measures of state fidelity, $\mathcal{F}(\rho_1,\rho_2) = \left( \textrm{Tr} \left( \sqrt{ \sqrt{\rho_2} \rho_1 \sqrt{\rho_2} }\right) \right)^2$
or alternatively the more generous measure $\sqrt{\mathcal{F}(\rho_1,\rho_2)}$. When applied to physically allowed states, both measures are non-negative and reach a maximum value of 1 when $\rho_1 = \rho_2$.
The fidelity used in the main text calculates $\mathcal{F}$ when comparing the gathered density matrix with the target density matrices. Examples of gathered versus ideal populations are shown in Figure \ref{skyscrapers}.

\section{Acknowledgements}
We gratefully acknowledge helpful discussions with A.~J.~Leggett, J.~Butterfield, G.~Milburn, D.~Loss, A.~Ardavan and V.~Watson. We thank EPSRC for supporting work at Oxford through EPSRC (EP/F028806/1), through CAESR (EP/D048559/1), and the Oxford-Keio collaboration through the JST-EPSRC SIC Program (EP/H025952/1). This work was supported by the National Research Foundation and Ministry of Education, Singapore, the Royal Society, the Clarendon Fund and St John's College Oxford.
\section{Author contributions}
G.C.K., S.S., E.M.G., J.J.L.M., G.A.D.B. and S.C.B. performed the theoretical analysis, designed the experiments, analysed the results and wrote the paper. S.S. and J.J.L.M. performed the experiments. H.R., N.V.A., P.B.. and H.-J.P. grew the $^{28}$Si crystal. K.M.I. and M.L.W.T. analysed and prepared the sample. 
\newpage

\renewcommand{\thesection}{S.\arabic{section}}
\renewcommand{\thesubsection}{\thesection.\arabic{subsection}}
 
\makeatletter 
\def\tagform@#1{\maketag@@@{(S\ignorespaces#1\unskip\@@italiccorr)}}
\makeatother
 
\renewcommand{\thefigure}{S\arabic{figure}}
\renewcommand{\figurename}{SUPPLEMENTARY FIGURE}
\renewcommand{\thetable} {S\arabic{table}} 
\renewcommand{\tablename}{SUPPLEMENTARY TABLE}
\newcommand{\citenumfmt}[1]{\textit{S#1}} 
\onecolumngrid
\appendix
\section{Supplementary Methods}

\subsection{Constraints on macrorealism}
Recall that each of the three core experiments are resolved into a further two sub-experiments, making six experiments in all. A macrorealist, under the assumption of non-invasive measurability, will concede that in the six experiments (which are performed on an identical initial state, and under the same conditions governing dynamics, i.e. the same Hamiltonian), the combined and post-selected results will be entirely equivalent to the family of three core experiments, each pair of circuits being equivalent to a single member of that family. Failure to post-select the results of measurements (as in e.g Ref.\cite{XuLiZou2011}) severely weakens the argument, and effectively introduces an extra assumption, namely that {\sc cnot} gates are always non-invasive. With proper post-selection then, the constraints (derived below) that are manifested in the Leggett-Garg inequality apply equally to the combined and post-selected results of the six lab experiments as they do to a single ideal experiment. This argument makes use of an additional assumption named `Induction'. This is an assumption about the behaviour of identically prepared and identically treated ensembles, and essentially states that causality only runs forwards in time~\cite{LeggettGarg2010}. We take this assumption as self evident and so do not state it explicitly in the main paper. Furthermore we believe that this assumption is equally required by experiments utilising a spatial ensemble and those using a time ensemble.

All macrorealist theories are required to predict measurement statistics for the correlators involved in the Leggett-Garg inequality. The underlying theory of macrorealism, if it is to be consistent, must abide by the conservation of probability and other consistency conditions. For example consider a general macrorealist theory assigning probabilities $\mathbb{P}(\updownarrow_1\updownarrow_2\updownarrow_3)$ to each possible evolution of the system:
\begin{table}[h!]
\begin{center}
\begin{tabular}{ccc|c|c}
$t_1$ & $t_2$ & $t_3$ &$\mathbb{P}$&$f_{LG}$\\ \hline
$\uparrow$& $\uparrow$ &$\uparrow$&$\mathbb{P}(\uparrow_1\uparrow_2\uparrow_3)$&4\\
$\uparrow$& $\uparrow$ &$\downarrow$&$\mathbb{P}(\uparrow_1\uparrow_2\downarrow_3)$&0\\
$\uparrow$& $\downarrow$ &$\uparrow$&$\mathbb{P}(\uparrow_1\downarrow_2\uparrow_3)$&0\\
$\uparrow$& $\downarrow$ &$\downarrow$&$\mathbb{P}(\uparrow_1\downarrow_2\downarrow_3)$&0\\
$\downarrow$& $\uparrow$ &$\uparrow$&$\mathbb{P}(\downarrow_1\uparrow_2\uparrow_3)$&0\\
$\downarrow$& $\downarrow$ &$\downarrow$&$\mathbb{P}(\downarrow_1\downarrow_2\downarrow_3)$&0\\
$\downarrow$&$\downarrow$&$\uparrow$&$\mathbb{P}(\downarrow_1\downarrow_2\uparrow_3)$&0\\
$\downarrow$& $\downarrow$ &$\downarrow$&$\mathbb{P}(\downarrow_1\downarrow_2\downarrow_3)$&4
\end{tabular}
\end{center}
\end{table}

\noindent For consistency we have 
\begin{align*}
\sum_{\updownarrow_1}\sum_{\updownarrow_2}\sum_{\updownarrow_3}\mathbb{P}(\updownarrow_1\updownarrow_2\updownarrow_3)=1,
\end{align*}
and for example
\begin{align*}
\mathbb{P}(\uparrow_1\downarrow_3)=\mathbb{P}(\uparrow_1\uparrow_2\downarrow_3)+\mathbb{P}(\uparrow_1\downarrow_2\downarrow_3).
\end{align*}
Using these conditions each correlator may be calculated from the macrorealist table by choosing the two appropriate rows for each two-time correlator (tracing out the column for whichever time is not needed), i.e. :
\begin{table}[h!]
\begin{center}
\begin{tabular}{cc|c|c}
$t_1$ & $t_2$ &$\mathbb{P}$&$Q(t_1)Q(t_2)$\\ \hline
$\downarrow$& $\downarrow$ &$\mathbb{P}(\downarrow_1\downarrow_2\uparrow_3)$+$\mathbb{P}(\downarrow_1\downarrow_2\downarrow_3)$&\phantom{-}1\\
$\downarrow$& $\uparrow$ &$\mathbb{P}(\downarrow_1\uparrow_2\uparrow_3)$+$\mathbb{P}(\downarrow_1\uparrow_2\downarrow_3)$&-1\\
$\uparrow$& $\downarrow$ &$\mathbb{P}(\uparrow_1\downarrow_2\uparrow_3)$+$\mathbb{P}(\uparrow_1\downarrow_2\downarrow_3)$&-1\\
$\uparrow$&$\uparrow$&$\mathbb{P}(\uparrow_1\uparrow_2\uparrow_3)$+$\mathbb{P}(\uparrow_1\uparrow_2\downarrow_3)$&\phantom{-}1
\end{tabular}
\end{center}
\end{table}

One then multiplies each pairwise sum of probabilities by $\pm1$ according to whether that row was a correlation or anti-correlation. The lower bound for the Leggett-Garg inequality arises from the frustration of a given state being anti-correlated with at most one of the other states (but not both), and the fact that because no single evolution of the system can violate the inequality, no statistical sampling will. 

\noindent Each of the classical trajectories can be probed non-invasively, by post-selecting populations from the appropriate circuit. An experimenter extracts correlations in the following way, with populations labelled $|\textrm{system}\rangle|\textrm{ancilla}\rangle$:
\begin{table}[h!]
\begin{center}
\begin{tabular}{cc|c|cc}
$t_i$ & $t_j$ & Population & Non Invasive?& Correlation?\\ \hline
$\downarrow$&$\downarrow$&$|\downarrow\downarrow\rangle$&{\sc CNOT}&+\\
$\downarrow$&$\uparrow$&$|\uparrow\downarrow\rangle$&{\sc CNOT}&-\\
$\uparrow$&$\downarrow$&$|\downarrow\downarrow\rangle$&{\sc anti-CNOT}&-\\
$\uparrow$&$\uparrow$&$|\uparrow\downarrow\rangle$&{\sc anti-CNOT}&+
\end{tabular}
\end{center}
\end{table}

\subsection{The stationarity assumption}
Although others have used it (sometimes implicitly), the additional assumption of stationarity is first given explicitly by Huelga et al. \cite{HuelgaMarshallSantos1995}:
\begin{quotation}
``\ldots the evolution from $t_1$ to $t_2$ is governed by the same stochastic differential equation as the evolution from $t_2$ to $t_3$, and this implies stationarity; that is $K(t_1,t_2)=K(t_1-t_2)$''.
\end{quotation}
This assumption is often used to redefine the Leggett-Garg Inequality
\begin{align}
f=K(\tau)+K(2\tau)\geq-1
\label{reLGI}
\end{align}
or similar. We note that there exist numerous macrorealist theories (which make predictions by distributing probability in the way outlined above) which are capable of violating \eqref{reLGI}. Consider a macrorealist theory which has $\theta$ as it's hidden variable, and flips from one of it's states to the other with a probability proportional to the cosine squared of this angle.  Such a theory is clearly capable of predicting Rabi oscillations. We take it to be an important feature of the original Leggett-Garg inequality that it is not violated by such theories. 

\subsection{Reducing the venality through hyperpolarisation}
The unitary nuclear rotation $U$ may be performed in a manner which is conditional on the system being in the `correct' ancilla state $\downarrow$ because the postselected data will always correspond to the unitary operation U having been applied.  If the rotation is conditional in this way, one of the two `bad' populations becomes inactive and will not experience any evolution whatsoever in the course of the protocol (specifically state $\ket{\downarrow\uparrow}$ for the {\sc cnot} circuits and $\ket{\uparrow\uparrow}$ for the anti-{\sc cnot} circuits). The inactive state does not participate in the experiment and may be ignored. By minimising the population of the \emph{single} active bad population we can reach a reduced effective venality. If the population distribution of all four energy levels is the same for the initial state of both circuits in each pair we have e.g. in the $\{\ket{\downarrow\downarrow},\ket{\downarrow\uparrow},\ket{\uparrow\downarrow},\ket{\uparrow\uparrow}\}$ basis 
\begin{align*}
\rho_C=\rho_A=
\frac{1}{Z}\left(
\begin{array}{cccc}
a &0   & 0 &0    \\
 0 & c  & 0 &0   \\
 0 &  0 &  b& 0 \\
0  &	0&0	&d
\end{array}
\right)
\end{align*}
where $\rho_C$ and $\rho_A$ are initial states prepared for {\sc cnot} and anti-{\sc cnot} circuits respectively, and $Z=a+b+c+d$ in both cases. The following expressions describe the lower bounds on quantum mechanical (QM), Moderate macrorealist (MMR) and Adversarial macrorealist (AMR) predictions: 
\begin{align*}
g_{QM}&\geq\phantom{-}\frac{1}{Z}  (a+b-c-d)(\textrm{cos}2\theta+2\textrm{cos}\theta) \\
g_{MMR}&\geq-\frac{1}{Z}(a+b)\\
g_{AMR}&\geq-\frac{1}{Z}(a+b+3c+3d)
\end{align*}
where $g=K_{12}+K_{13}+K_{23}$ and $f=g+1$. The venality $\zeta=(c+d)/Z$ allows one to write
\begin{align}
g_{QM}&\geq\phantom{-}(1-2\zeta)(\textrm{cos}2\theta+2\textrm{cos}\theta)\label{gq}\\
g_{MMR}&\geq-(1-\zeta)\label{gm}\\
g_{AMR}&\geq-(1-\zeta)-3\zeta.
\label{ga}
\end{align}
In thermal equilibrium $(a,b,c,d)=(1,\alpha,1,\alpha)$ and so in general $\zeta=2\alpha/(2+2\alpha)$. When oscillations are only driven on those primary systems which were paired with a correctly initialised ancilla, one (system,ancilla) state always remains unused throughout the experiment. We exploit this fact by hyperpolarising the system so that the remaining active state has a lower population than is possible in thermal equilibrium at a given temperature. If the population distribution is identical across only the three active levels of the experiment we have
\begin{align*}
\rho_C=
\frac{1}{Z}\left(
\begin{array}{cccc}
a  &0   & 0 &0    \\
 0 & [c]  & 0 &0   \\
 0 &  0 &  b& 0 \\
0  &	0&0	&d
\end{array}
\right)
\end{align*}
for the {\sc cnot} circuit and
\begin{align*}
\rho_A=
\frac{1}{Z}\left(
\begin{array}{cccc}
a  &0   & 0 &0    \\
 0 & d  & 0 &0   \\
 0 &  0 &  b& 0 \\
0  &	0&0	&[c]
\end{array}
\right)
\end{align*}
for the anti-{\sc cnot} circuit with $Z=a+b+c+d$ as usual. The inactive state is denoted with $[\phantom{x}]$. These different initial states, although physically distinct, are logically identical because the relevant active energy levels have the same population distribution. The predictions are now
\begin{align*}
g_{QM}&\geq\phantom{-}\frac{1}{Z}  (a+b-2d)(\textrm{cos}2\theta+2\textrm{cos}\theta)\\
g_{MMR}&\geq-\frac{1}{Z}(a+b)\\
g_{AMR}&\geq-\frac{1}{Z}(a+b+6d).
\end{align*}
Note that all predictions are independent of the inactive state with population $c$, except for in the normalisation $Z$. The normalisation can be arbitrarily scaled without affecting the comparison of the three predictions for $g$ (or for $f$) since they will all be affected linearly in the same fashion. We choose to multiply $g$ by $Z/(a+b+2d)$ so that there is a normalisation of $a+b+2d=Z_r$ and no longer any dependence on $c$. This allows us to define the venality as $\zeta=2d/Z_r$ 
and to recover equations \eqref{gq},\eqref{gm},\eqref{ga}. This technique is equivalent to supplying the single four level population distribution 
\begin{align*}
\rho^\prime=
\frac{1}{Z_r}\left(
\begin{array}{cccc}
a  &0   & 0 &0    \\
 0 & d  & 0 &0   \\
 0 &  0 &  b& 0 \\
0  &	0&0	&d
\end{array}
\right)
\end{align*}
to both types of circuit.  Using hyperpolarisation we achieve $(a,b,c,d)=(1,\alpha,\alpha,\alpha^2)$ so that $\zeta=2\alpha^2/(1+\alpha+2\alpha^2)$. 
\subsection{Effect of detuned pulses}
In the ideal scenario, the experimenter applies either the {\sc cnot} or anti-{\sc cnot} to the primary system-ancilla pair to perform the non-invasive measurement. In real spin resonance experiments each of the pulses will excite finite amplitude in the unwanted transition (i.e. it is not infinitely far off resonance). The post-selection procedure will remove any pairs from the ensemble which are affected by a microwave pulse, detuned or not;  but of course this post-selection is ill-informed for those pairs in which the ancilla is incorrectly initialised. To allow for this one can simply expand the venality to include a fraction $\Delta$ of the inactive state population. Note that this $\Delta$ can be arbitrarily minimised in spin-resonance experiments by for example increasing the duration of the pulses which are applied, or using a sample with a larger splitting between the two microwave frequencies. In our experiment the $\Delta$ is less than 0.04 and we have confirmed that the corresponding correction to venality makes little difference to the  degree of violation of our Leggett-Garg inequality.

Note that it is also important that the physical implementation of the {\sc cnot} (and anti-{\sc cnot}) operations is such that the primary system receives no perturbation when it is in state $\downarrow$; it would not be acceptable to implement the {\sc cnot} as a series of low level operations, some of which perturb the primary system: even if their net effect is that of the {\sc cnot} (as is the case for example with a controlled phase gate plus single qubit rotations).
\section{Supplementary References}

\end{document}